\documentclass[12pt]{article}
\usepackage[utf8]{inputenc}
\usepackage[letterpaper, margin=1in]{geometry}  
\usepackage{titlesec}  
\usepackage{graphicx}  
\usepackage{amsmath, amssymb}   
\usepackage{lipsum}    
\usepackage{hyperref}  
\usepackage{enumerate} 
\usepackage{fancyhdr}  
\usepackage{braket}   
\usepackage{cleveref}  
\usepackage{tikz-cd}
\usepackage[english]{babel} 
\usepackage{tikz}
\usetikzlibrary{babel} 
\usetikzlibrary{positioning} 
\usetikzlibrary{calc}
\usepackage{setspace}
\usepackage[numbers]{natbib}
\usepackage{tikz}
\titleformat{\section}{\large\bfseries\raggedright}{\thesection.}{0.5em}{}
\titleformat{\subsection}{\normalsize\bfseries\raggedright}{\thesubsection.}{0.5em}{}

\begin{document}

\begin{center}
    \begin{spacing}{1.3} 
        \textbf{\ Relational Quantum Dynamics (RQD): An Informational Ontology}\\[20pt]
        \large Arash Zaghi$^{\dagger}$ \\[5pt]
    \end{spacing}
\end{center}

\footnotetext{$^{\dagger}$ ORCID ID: \href{https://orcid.org/0000-0003-2246-2911}{0000-0003-2246-2911}, Email: \texttt{arash.esmaili\_zaghi@uconn.edu}}

\begin{center}
    \textbf{Abstract}\\[10pt]  
    \begin{minipage}{0.8\textwidth}
        \small \
        Quantum mechanics has transformed our understanding of reality, yet deep philosophical puzzles remain unresolved. Is there a consistent way to describe quantum measurement, the emergence of space and time, and the role of observers within one coherent ontology? This paper introduces Relational Quantum Dynamics (RQD), an interpretation that places relationships and information—not isolated objects—at the foundation of physics. Unlike traditional views, such as the Copenhagen interpretation, Everett’s many-worlds theory, or hidden-variable approaches, RQD denies a single observer-independent reality, instead proposing that facts only become definite within specific observational contexts. It explicitly addresses foundational no-go results, including Bell’s theorem, the Kochen–Specker theorem, and the Frauchiger–Renner paradox, by replacing absolute facts with relational, context-dependent truths. RQD rests on five interlocking principles: (1) quantum states are contextual and observer-relative, (2) time emerges from quantum interactions rather than existing as an external parameter, (3) space is constructed from patterns of quantum entanglement, (4) observers emerge as systems with highly integrated information, and (5) classical reality appears through quantum decoherence without special wavefunction collapse. Philosophically, RQD aligns with ontic structural realism, proposing that the fundamental ontology of the universe consists entirely of informational relations. It avoids collapsing into relativism or solipsism by ensuring intersubjective agreement through physical interactions. Ultimately, RQD may offer a unified picture in which quantum mechanics, spacetime, and observers are no longer separate domains but aspects of one interconnected informational web. 

    \end{minipage}
\end{center}

\begin{center}

\end{center}

\section{Introduction}
\textbf{Research Question:} \textit{Can one consistently describe quantum measurement, spacetime emergence, and observers within a single ontology?} This question lies at the heart of modern physics \cite{Gambini2015,huggett2021out}. Quantum mechanics, in its standard form, treats measurement outcomes as fundamentally contextual or observer-linked (the infamous Schrödinger’s cat or Wigner’s friend scenarios)\cite{Wigner1961,Proietti2019,Frauchiger2018}, whereas general relativity treats spacetime as an objective, observer-independent continuum\cite{taylor1992spacetime}. These frameworks lead to deep tensions. For example, quantum theory suggests an observer’s choice can nonlocally affect outcomes (violating classical intuitions of locality and realism), while relativity implies that time and space evolve without reference to any observer\cite{bancal2012quantum,banerjee2016quantum}. How can \textit{observers}, who are themselves physical systems, be consistently included in quantum descriptions? Is \textit{spacetime} perhaps not fundamental at all, but emergent from quantum relationships? Such questions motivate our investigation. We seek a framework in which quantum measurement, the emergence of spacetime structure, and the role of observers are \textit{all} accounted for within a single coherent ontology.

Recent developments provide clues that \textit{information} and \textit{relations} could serve as the unifying thread. In quantum foundations, \textit{relational} interpretations hold that the quantum state of a system is meaningful only relative to some observer or reference system \cite{Rovelli1996,sep-qm-relational,nicolaidis2012relational,ananthaswamy2018frauchiger}. In quantum gravity, there are hints that spacetime geometry itself might arise from patterns of quantum entanglement \cite{VanRaamsdonk2010,RyuTakayanagi2006,Swingle2009Entanglement,swingle2018spacetime}. The \textit{problem of time} in quantum gravity has prompted ideas like the thermal time hypothesis, which posits that time flow might be an emergent, state-dependent property rather than an absolute external parameter \cite{connes1994von,chua2024time}. Meanwhile, research on decoherence has illuminated how classical measurement outcomes can emerge from quantum dynamics without introducing any mystical collapse \cite{zurek1991decoherence,joos1985emergence,zurek2003decoherence}. Even theories of consciousness, such as Integrated Information Theory, quantify how strongly a system’s information is unified, potentially offering a physical criterion for when a collection of quantum particles forms a single “observer” with definite experiences \cite{Tononi2004,Tononi2016IntegratedInformationTheory,mediano2020measuring}. These insights remain largely disjoint in existing literature. Each addresses a piece of the puzzle—quantum measurement, time, space, information, or observation—but no single framework synthesizes them. As a result, foundational paradoxes persist. For instance, the Frauchiger–Renner no-go theorem shows that quantum theory, as usually understood, cannot consistently describe observers who themselves use quantum theory \cite{Frauchiger2018,ananthaswamy2018frauchiger}. Likewise, Wigner’s friend thought experiments highlight contradictions if one assumes a \textit{single, observer-independent set of facts} in a quantum scenario \cite{Brukner2018}.

In this paper, we propose \textbf{Relational Quantum Dynamics (RQD)} as a unified interpretative ontology that tackles these issues head-on. RQD is built on the idea that \textit{relations} and \textit{information}, rather than isolated objects or absolute spacetime, form the bedrock of physical reality. By treating every physical interaction as an exchange of information between systems (without privileging any “classical” observer), RQD aims to reconcile quantum mechanics with an emergent spacetime and to include observers \textit{within} the physics, not outside it. In brief, our thesis is that a carefully constructed relational, information-theoretic ontology can consistently describe quantum measurements, the emergence of time and space, and the role of observers \textit{all together}. We will show that RQD relaxes certain orthodox assumptions, such as observer-independent state descriptions and the existence of an external global time, while upholding the empirical core of quantum theory. In doing so, it provides a new way to resolve or \textit{dissolve} the notorious paradoxes and no-go theorems that plague conventional interpretations.

\textbf{Overview:} We begin by situating RQD within the landscape of quantum interpretation (Section 1), contrasting it with standard approaches (Copenhagen, Everett, Bohmian mechanics, objective collapse, etc.) and identifying which traditional assumptions it abandons. In Section 2, we examine four key no-go theorems or paradoxes (Bell’s theorem, the Kochen–Specker theorem, Wigner’s friend, and Frauchiger–Renner), and identify the philosophical tensions they highlight (for example, locality vs. realism, context-dependence, objectivity of facts). These will set the stage for the demands on any successful ontology. Section 3 introduces the five foundational principles (or “pillars”) of RQD—\textbf{(1)} contextual quantum states, \textbf{(2)} modular time, \textbf{(3)} entanglement-built geometry, \textbf{(4)} integrated information, and \textbf{(5)} decoherence—formulated as axioms with conceptual and philosophical justification. In Section 4, we discuss the philosophical implications of RQD, align it with ontic structural realism and perspectival realism, and compare it with QBism and other views. We also address potential objections, including concerns about empirical underdetermination, testability, and whether RQD lapses into panpsychism or idealism. Section 5 concludes with a summary of the philosophical benefits of RQD and outlines future directions for research at the intersection of physics and philosophy.

\section{RQD in the Landscape of Quantum Interpretations}
\textbf{Standard Interpretations:} A wide spectrum of interpretations of quantum mechanics exists, each with its own ontology and set of assumptions. At one end, the orthodox \textit{Copenhagen} approach insists on a duality between quantum and classical: quantum systems evolve unitarily until a measurement causes an abrupt, observer-effected wavefunction collapse \cite{walker2024copenhagen,zinkernagel2016bohr}. Copenhagen pragmatically avoids describing observers quantum-mechanically by positing a classical realm, but at the cost of an unresolved “Heisenberg cut” (an arbitrary division between observer and system). At another extreme, the \textit{Everettian} or \textit{Many-Worlds} interpretation assumes the wavefunction never collapses; every possible outcome occurs in an ever-branching multiverse. Everett restores unitarity at the global level but demands a single, observer-independent wavefunction for the universe, raising the question of how \textit{our} particular experience of a definite outcome arises (the basis and probability problems) \cite{lazarovici2023everett,sebens2016self}. \textit{Hidden-variable theories} like \textit{Bohmian mechanics} keep a single reality but augment quantum theory with additional ontic variables (for example, particle positions guided by the wavefunction) to ensure definite outcomes \cite{singh2008bohm,lazarovici2018observables}. These theories, however, assume an absolute space and time background and a privileged reference frame for the guiding equation, and they must accommodate observable violations of Bell’s inequalities by abandoning locality \cite{berndl1996epr}. \textit{Objective collapse theories}, like GRW and variants, modify the Schrödinger dynamics with stochastic physical collapses, invoking new dynamics to solve the measurement problem at the expense of introducing spontaneity or new forces \cite{ghirardi1986unified}. Finally, \textit{epistemic interpretations}, like \textit{QBism}, radically personalize the quantum state, treating it not as an objective entity at all but as an agent’s credence or information about outcomes \cite{Pienaar2021,fuchs2016participatory,fuchs2013quantum}. Such interpretations avoid any literal wavefunction collapse or branching at the ontology level, but they raise the question of whether there is any mind-independent reality described by quantum theory.

\textbf{Position of RQD:} Relational Quantum Dynamics can be viewed as a new entrant in this landscape, sharing characteristics with some interpretations while diverging on key points. RQD explicitly \textit{abandons the notion of an observer-independent quantum state}, aligning with the spirit of Rovelli’s \textit{Relational QM} and with QBism’s insistence on the role of the agent’s perspective \cite{Pienaar2021}. In RQD, there is no God’s-eye-view wavefunction for the whole universe that all observers must agree upon. Instead, the quantum state is \textit{indexical}, it is always the state of system S relative to system O. By relaxing the assumption of a single absolute state, RQD removes the need for an externally defined “collapse” or a many-worlds proliferation of objective branches. It also dodges the inconsistency that arises when one tries to have two observers (Wigner and his friend, for example) both apply quantum theory to each other and arrive at a single, unified account. Each observer in RQD has their own valid state assignment for a system, and these need not be globally reconciled into one “true” state independent of perspective.

At the same time, RQD maintains a \textit{realist} spirit. It is not merely an instrumental or subjective account. It differs from QBism by holding that quantum states are informational realities, not just personal beliefs. RQD’s ontology is one of \textit{relations} and \textit{information}, which exist objectively (in the sense of being \textit{there} in the world) but only as networked, perspective-dependent quantities. In philosophical terms, it leans toward \textit{perspectival realism}: there is a mind-independent world, but facts in that world are relativized to particular informational relationships (observers and systems). This move relinquishes the classical ideal of observer-independent properties, but it strives to keep science from sliding into solipsism by explaining how different observers can eventually agree on shared facts through communication and interaction (we will expand on this in Section 4).

Another core assumption RQD challenges is the existence of a single external \textit{time parameter}. In standard quantum mechanics, time is an absolute background variable (a remnant of classical Newtonian thinking) not subject to quantum treatment. But attempts to quantize gravity suggest that there is no global time at the most fundamental level. The Wheeler–DeWitt equation famously yields a “frozen” universe state \cite{rotondo2022wheeler}. RQD, influenced by ideas like the thermal time hypothesis, treats time as an \textit{emergent phenomenon} \cite{connes1994von} tied to the state of an informational subsystem. By doing so, RQD relaxes the assumption of an external classical clock. This helps resolve the “problem of time” and allows RQD to integrate quantum mechanics with a dynamical spacetime—time and space become properties that emerge from quantum relations rather than pre-existing stages on which physics unfolds.

In summary, RQD stakes out a unique position: it is a \textit{fully quantum, non-dualistic interpretation} (no split between quantum and classical regimes; observers and measuring devices are just quantum systems). It \textit{forgoes observer-independent states and absolute time}, yet \textit{retains realism} by positing an underlying quantum-informational structure that constitutes reality. In doing so, RQD must confront the same empirical constraints as other interpretations. We now turn to those constraints: the famous no-go theorems and paradoxes that any satisfactory interpretation must address. How does RQD navigate the trade-offs between realism, locality, objectivity, and consistency highlighted by these results?

\section{No-Go Theorems and Philosophical Tensions}
Foundational results in quantum theory reveal which classical intuitions cannot hold in our world. Each no-go theorem pinpoints a clash of principles, forcing us to give up something (be it determinacy, locality, a single reality, etc.). We briefly review four key cases and identify the philosophical tension each illustrates, setting the stage for RQD’s resolutions:

\begin{itemize}
    \item \textbf{Bell’s Theorem:} \textit{Locality vs. Reality}. Bell’s theorem proves that \textit{no local hidden-variable theory can reproduce all the predictions of quantum mechanics} \cite{Aspect1982bell,Bell1964epr}. Empirically, entangled particles violate Bell’s inequality, so we must abandon either the principle of locality or the idea that quantum properties have pre-determined real values (or both). This challenges a classical realist worldview where objects have properties independent of distant events. Philosophically, Bell’s result forces a choice between \textit{local realism} and the nonlocal, holistic nature of quantum correlations. Most physicists accept that nature is nonlocal in the limited sense allowed by relativity (no \textit{signaling} faster than light, but entangled correlations exist without mediating signals). \textbf{RQD’s stance:} The world is holistic and relational; quantum correlations are \textit{not} built from separate local pieces but are inherent in the structure of reality. RQD thus forgoes the classical assumption of separable, independently real parts. However, because each quantum interaction in RQD is still constrained by relativistic signaling (information transfer requires causal contact), \textit{no observer in RQD ever sees a violation of relativistic causality}. The nonlocality is epistemic unless and until observers compare notes by coming together and exchanging information classically. In short, Bell’s theorem is not paradoxical in RQD because RQD does not assume counterfactual definiteness or independent local reality in the first place. In RQD, the relational web of information is the reality, and it is globally holistic.
    \item \textbf{Kochen–Specker:} \textit{Contextuality vs. Value Definiteness}. The Kochen–Specker theorem (1967) showed that it is \textit{impossible to assign definite values to all quantum observables in a non-contextual way} \cite{Kochen1967problem}. In other words, one cannot presume that each quantum property has a pre-existing value that does not depend on how you measure it (the measuring context). This result undercuts the classical idea of \textit{objectivity of properties}, that physical quantities have observer-independent values. The philosophical tension here is between \textit{realism about intrinsic properties} versus \textit{contextuality} \cite{Budroni2022}, which says that measurement context helps define what properties can be said to have values). \textbf{RQD’s stance:} Embrace \textit{contextuality} as fundamental. In RQD, a property value (an outcome) is meaningful only relative to an interaction context; effectively, relative to an observer-system relation. There is no “view from nowhere” that assigns values to all observables at once. By taking this perspectival view, Kochen–Specker is not a problem but an affirmation that quantum truths are \textit{perspectival}. RQD reframes the notion of a “fact” to “X has value v \textit{for observer O}”, rather than “X has value v \textit{absolutely}.” This way, the would-be Kochen–Specker contradiction evaporates; a value map assigning definiteness to all observables at once is neither required nor possible. What classical thinking calls “objectively real values” are in RQD replaced by \textit{relation-dependent facts}. Consistently, no single overarching assignment of values exists that would violate the Kochen–Specker constraints.
    \item \textbf{Wigner’s Friend:} \textit{Observer-Independent Facts vs. Relational Facts}. The Wigner’s friend thought experiment imagines an observer’s friend making a quantum measurement inside an isolated lab, while Wigner outside treats the entire lab (friend included) as a quantum system \cite{Wigner1961,Wigner1995,Bong2020}. The friend obtains a definite result, yet Wigner, using quantum theory, would describe the friend+system in a superposition. Classical intuition demands that either the friend’s result was not “real” or that Wigner’s state description is wrong. This leads to a paradox of \textit{objectivity}: can there be a single, observer-independent account of what “happened”? Recent extensions by Frauchiger and Renner make this even sharper, suggesting that no single-world narrative can accommodate all observers’ predictions consistently \cite{butterfield1999emergence}. The philosophical tension is whether \textit{facts} are absolute or \textit{observer-relative}. \textbf{RQD’s stance:} There is \textit{no single, absolute narrative} in such situations. Each observer has a valid description within their own frame of reference, and there is \textit{no fact of the matter} about the outcome \textit{until} one brings the observers together. RQD fully embraces the idea that “facts are indexical”. In the Wigner’s friend scenario, for the friend inside, “the experiment yielded outcome $O$” is a fact; for Wigner outside, the lab is in a superposed state and that same statement is not a fact (for him) until he interacts with the lab. Importantly, RQD allows \textit{both} of these descriptions to coexist without conflict because they are descriptions \textit{from different viewpoints}. There is no logical contradiction because the statements “Outcome is $O$” and “Outcome is not definitively $O$” are \textit{each indexed to different observers}. When Wigner eventually interacts with (measures) the lab, standard quantum theory (plus decoherence) ensures he will find a definite record that aligns with the friend’s recorded outcome, thereby \textit{relating} the two perspectives and restoring consistency. RQD thus resolves the Wigner’s friend paradox by denying a global, observer-independent set of facts---a move sometimes called giving up “single reality”. What is sacrificed is \textit{objectivity} in the naive sense; what is retained is consistency (no one observer ever sees a violation of quantum mechanics) and \textit{intersubjective agreement} once communication occurs. Different agents’ accounts are not unified by an external god’s-eye perspective, but they \textit{become compatible through interaction}: when observers compare notes, their relational facts \textit{align}. This is ensured by decoherence and the structure of quantum interactions, as we will discuss.
    \item \textbf{Frauchiger–Renner No-Go:} \textit{Single World vs. Consistency}. Frauchiger and Renner (2018) \cite{Frauchiger2018} considered an elaborated Wigner’s friend setup and proved a no-go theorem: \textit{no interpretation of quantum mechanics that insists on a single, observer-independent reality can consistently account for the predictions of all agents} \cite{dorato2017defense,ananthaswamy2018frauchiger}. In effect, they showed that assuming (Q) quantum theory is universally valid, (C) logic is consistent, and (S) a single definite outcome occurs for each experiment leads to a contradiction. This theorem crystallizes the incompatibility between \textit{universal validity} of quantum mechanics and a \textit{“one-world” assumption}. Philosophically, it forces a hard choice: we might abandon the idea of a single classical reality, as Everett does by going many-worlds, or abandon universal applicability, as Copenhagen does by positing a classical domain, or even modify quantum theory. \textbf{RQD’s stance:} RQD takes the Frauchiger–Renner result as validation of its core idea – the assumption of a single, observer-independent reality is what has to give. Indeed Frauchiger and Renner themselves note that interpretations like relational QM or many-worlds evade the contradiction by rejecting a single global reality. RQD does so in a \textit{relational} (not many-worlds) manner: it jettisons the notion of a singular truth accessible to all observers, and in return it gains internal consistency. Different observers’ “worlds” (sets of facts) are \textit{complementary but not simultaneously co-real}. Only when interactions bring those observers together can their knowledge be compared, at which point standard quantum rules ensure they agree on the shared events. By dropping the \textit{Single World} assumption (S) and carefully specifying how cross-perspective consistency is achieved via interactions, RQD escapes the Frauchiger–Renner no-go theorem. In summary, \textit{objectivity} in the strict sense is traded for \textit{relational consistency}. This is a move that is philosophically radical (truth becomes perspectival) but, we argue, a necessary one to avoid the contradictions laid bare by these no-go theorems.
\end{itemize}

Having identified these pressures from fundamental theorems, we see a common thread: quantum theory seems to demand a \textit{relational, context-dependent notion of reality}. RQD picks up this thread and weaves it into a positive ontology. Rather than adding arbitrary mechanisms or shying away from quantum description of big systems, RQD modifies the \textit{conceptual foundations}: it posits that only a network of relative facts exists and that classical reality, including spacetime and definite outcomes, emerges from deeper quantum-informational relations. We now formalize this vision by laying out the five key principles of RQD, each addressing one facet of the quantum reality puzzle.

\section{Five Principles of Relational Quantum Dynamics}
We propose five interlocking principles as the foundation of RQD. They can be viewed as \textit{axioms} or postulates that jointly define the ontology. Each principle is motivated by prior theoretical ideas (in quantum foundations, quantum gravity, or information theory), but \textit{the novelty of RQD lies in combining all five} in one framework and insisting that \textit{omitting any one} reopens paradoxes. We enumerate and explain each principle along with its conceptual and philosophical basis:

\begin{enumerate}
    \item \textbf{Contextual Quantum States.} \textit{Quantum states are relational and context-dependent, defined only with respect to an “observer” system.} In RQD, there is no absolute, observer-independent quantum state of a system. This principle follows the insight of Rovelli’s relational QM: the state, and values of physical quantities, refer to the relation between two systems, typically an observed and an observer. Different observers can give different yet valid accounts of the same event. By adopting this, RQD naturally circumvents paradoxes like Wigner’s friend. Each observer, or reference frame, has their own facts without the need for a privileged “God’s-eye view” to reconcile them. Philosophically, this is a shift to \textit{perspectival realism}. It asserts that reality consists of many partial viewpoints, none of which is \textit{the} view from nowhere. The \textit{metaphysical commitment} here is that what exists is a network of information exchanges (relations) rather than isolated objects with intrinsic properties. This aligns with ontic \textit{structural realism}, which holds that relations (structure) are primary and objects are at most nodes in that structure \cite{alai2017debates}. By embracing relational states, RQD gives up the classical ideal of observer-independent \textit{truth values}, but gains a consistent way to include observers \textit{within} the theory. As a payoff, the Frauchiger–Renner inconsistency is avoided because RQD denies the assumption that all observers’ statements can be combined into one single narrative.
    \item \textbf{Modular Time (Emergent Temporal Context).} \textit{Time is not a universal background parameter; rather, each subsystem can have its own internal time emergent from its state, for example, via entropy or correlations.} RQD posits that “time” is a derived concept, arising from the dynamics of information. This principle draws on the thermal time hypothesis of Connes and Rovelli and related ideas like Page and Wootters’ mechanism where entanglement can produce an internal clock \cite{connes1994von,wootters1982single}. The essence is that \textit{temporal ordering is context-dependent}: within a given system (or observational context), one can define a time flow, for example, the system’s entropic change provides a clock, but there is no global \$t\$ ticking for the universe. Philosophically, this undermines Newtonian absolutism about time and resonates with relational and Machian ideas (time as an aspect of relationships among objects) \cite{earman1989world}. RQD’s modular time addresses the “problem of time” in quantum cosmology by saying that what we call time is a higher-level, emergent phenomenon, a byproduct of entangled relationships and information thermodynamics. By dropping the assumption of an external time axis, RQD aligns with general relativity’s lesson that time is not fundamental. It also means that different subsystems can experience time at potentially different rates or have only approximate synchrony, much as general relativity taught us with gravitational time dilation. However, here time arises from quantum informational structure rather than gravity per se. This principle ensures that RQD can integrate a quantum description of the whole universe, which is “timeless” in the Wheeler–DeWitt sense, with the fact that observers inside the universe experience a flow of time. Time in RQD is an \textit{indexical} notion: “time for \textit{X}” is what emerges from \textit{X’s} state and interactions. This move has profound philosophical implications: it suggests that \textit{becoming} (the flow of time) is not an objective global feature but a perspectival aspect of subsets of the universe. The nature of temporal reality becomes akin to a context-dependent feature, potentially illuminating discussions by philosophers of time. For example, refer to arguments by Butterfield or Earman on whether time is emergent or fundamental \cite{butterfield1999emergence}.
    \item \textbf{Entanglement-Based Geometry.} \textit{Spatial relationships are constructed from entanglement, quantum entanglement patterns give rise to effective spatial connectivity and geometry.} This principle asserts that spacetime is not a fundamental stage but an emergent web woven by quantum correlations. It is inspired by results in quantum gravity and holography, especially the work of Van Raamsdonk showing that reducing entanglement between two regions causes those regions to “split apart” in the emergent geometry \cite{VanRaamsdonk2010,swingle2012constructing,swingle2018spacetime}. In RQD, \textit{distance and space} are interpreted as derived from the degree of entanglement: highly entangled systems effectively sit “close” in space (possibly connected by something like a wormhole in holographic scenarios), whereas unentangled or disentangled systems are “far” apart or in disconnected regions of space. Thus, as a system’s entanglement with others changes, the spatial picture changes, \textit{space itself is relational}. Philosophically, this is a form of \textit{ontic structural realism} about spacetime: the structure of entanglement (a graph or network of relations) is more fundamental than spacetime points. The “metric” properties (like distance) emerge from the network’s properties, for example, entanglement entropy might relate to spatial volume or area \cite{cao2016space}. This principle provides a mechanism for unifying quantum non-locality with classical locality: classical space emerges as a coarse-grained, approximate concept when quantum connections are rich enough to form a smooth fabric. When entanglement is lost beyond some scale, spacetime can tear or split, reflecting how in quantum gravity, spacetime can in principle change topology or connectivity. By including entanglement-based geometry, RQD links itself to ongoing debates about spacetime emergence in quantum gravity, suggesting that what we perceive as continuous space is an epiphenomenon of underlying quantum information links. This addresses the question: \textit{“Is spacetime built from something deeper?”}. RQD answers \textit{yes, from entanglement relations}.
    \item \textbf{Integrated Information and Observers.} \textit{When is a collection of particles an “observer” or a unified system? RQD answers: when it has a high degree of integrated information}. Borrowing from \textit{Integrated Information Theory (IIT)}, originally a theory of consciousness \cite{Tononi2004,Tononi2016IntegratedInformationTheory,Tononi2008IIT-ProvisionalManifesto,lewis2023intriguing,DURHAM2023PUBLISHED}, RQD incorporates the idea that a system which generates significantly more information as a whole than the sum of information generated by its parts can be considered a unified entity. In physical terms, such a system has strong interdependencies: its parts are so interrelated that the state of the whole cannot be decomposed without loss of information. RQD uses this as an \textit{objective criterion} for identifying stable “observers” or quasi-classical apparatuses within the quantum world. For example, a measuring device that has many internal correlations (record states) tying it together will have high integrated information, and thus it behaves as a single entity that can hold a definite record. Philosophically, this principle adds a \textit{neo-Aristotelian} twist: a whole is \textit{more than the sum of its parts} \cite{koslicki2006neo}, and that “more” (the surplus information generated by holistic structure) is taken as ontologically significant. It grounds the \textit{emergence} of higher-level individuals, like a conscious mind, or a functioning detector, in information-theoretic terms. By doing so, RQD avoids an arbitrary Heisenberg cut: there is no need to assume humans or macroscopically large objects are classical by fiat. Instead, whether something can be treated as an observer with definite perspective is a matter of degree, quantified by integrated information (often denoted $\Phi$). A simple system (low $\Phi$) does not have a single integrated perspective; a complex brain (high $\Phi$) does. This has an interesting philosophical consequence: it suggests a continuum between simple physical systems and complex observers, differing in degree of organizational complexity rather than a binary split. It thus provides a potential bridge in the \textit{mind–body problem}: consciousness (subjective experience) correlates with high integrated information, but that property is still a physical, quantitative one. RQD’s ontology thereby leans toward an \textit{informational monism} \cite{sevo2023informational}: it asserts that what fundamentally exists is information structure, and that at a certain level of integrated complexity, that same stuff manifests as what we call conscious experience. More on this in Section 4 when we discuss the metaphysical implications.
    \item \textbf{Decoherence and Classical Records.} \textit{Quantum dynamics plus decoherence suffice to explain the emergence of classical outcomes and records, without adding collapses.} Decoherence is the process by which a quantum system’s coherence (phase relations between states) is rapidly degraded due to interaction with its environment \cite{joos1985emergence}. The environment in effect “measures” the system incessantly, in random bases, which has the result of suppressing interference and selecting a stable set of \textit{pointer states} that can persist. RQD embraces decoherence as a key part of the story: it is how the multiplicity of possible quantum outcomes reduces, in each observer’s frame, to an effective single outcome (a stable record). Importantly, RQD \textit{does not posit wavefunction collapse as a fundamental physical law}; instead, apparent collapse is an emergent consequence of decoherence plus the relational perspective. From any one observer’s viewpoint, once entanglement with an environment has spread, and especially if the information has become redundant in the environment\cite{zwolak2016amplification}, the system of interest \textit{can be treated as if} it has a definite state; in other words the probability distribution has narrowed to a delta for all practical purposes. Another observer might be in a superposition relative to that basis, but by Principle 1, that’s fine, their facts differ until interaction brings them into alignment. Philosophically, this principle reaffirms a \textit{form of realism about records}: stable classical facts exist, but they are relative facts stabilized by decoherence. It resolves the measurement problem without mystery: Schrödinger’s cat is not “consciously observed” into life or death, but rather, decoherence (from air molecules, photons, etc.) in the cat’s environment ensures that in any given frame (such as the cat’s own cells, or a Geiger counter) the superposition of alive/dead \textit{branches into effectively distinct non-interfering outcomes} long before a human opens the box. Each branch is a relational world relative to some observer; no interference means they do not recombine. Thus, every observer sees a definite cat, and their interactions will confirm the cat is either alive or dead, never a superposition. Decoherence supplies the mechanism by which the \textit{quantum-to-classical transition} occurs dynamically, consistent with unitarity. RQD holds that once we accept the other principles, relational states, emergent time, entanglement geometry, integrated information, decoherence is the final piece that makes classical reality (complete with shared records and histories) emerge in each context. By not introducing any ad hoc collapse, RQD stays fully within quantum theory while still explaining why we \textit{experience} a classical world.
\end{enumerate}

These five principles form a mutually strengthening set. Each addresses a facet of the measurement and reality puzzle, and notably, each also corresponds to one of the tensions highlighted by the no-go theorems. For instance, \textit{Contextual States} directly answer Kochen–Specker and Wigner’s friend by denying universal value assignments or facts; \textit{Modular Time} speaks to the problem of time (giving up absolute time consistency); \textit{Entanglement-Based Geometry} links to nonlocality by accepting holistic connections (thus dovetailing with Bell’s insights); \textit{Integrated Information} gives a criterion for the Heisenberg cut (who or what is a “measurer”) without external collapse, thus speaking to the measurement problem; and \textit{Decoherence} quantitatively accounts for the emergence of classicality from quantum law (addressing the transition from quantum possibilities to effective single outcomes).

It is crucial to note that \textit{RQD asserts all five principles together}. If one omits any single pillar, the coherence of the framework breaks down, and familiar paradoxes re-emerge. For example, without contextual relational states, we would be forced back to a single global wavefunction and immediately face contradictions such as the Frauchiger–Renner paradox. Without treating time as emergent, a global clash with the static block universe of general relativity reappears. Without entanglement-induced geometry, we lose any account of how classical spacetime arises from quantum physics. Without integrated information, we have no physical handle on why certain systems (observers) have definite perspectives and others do not, risking either panpsychism (every particle “observes”) or an arbitrary classical/quantum divide. And without decoherence, we have no explanation for why superpositions turn into the familiar facts of our experience. In RQD, each principle “patches a hole” that the others alone cannot fully fix; together they form a logically tight package. This holistic structure is both a strength and a challenge: it means RQD is \textit{ambitious} in scope, aiming not just to interpret quantum mechanics in the lab, but to propose a unified ontology for quantum physics \textit{and} spacetime \textit{and} information. We acknowledge that each piece comes with open questions. For example, how exactly to quantify integrated information in quantum field terms, or how to formally derive geometry from entanglement in general cases. Nonetheless, the coherence of the overall picture is what we wish to emphasize. RQD offers a plausible sketch of reality in which the long-standing dualisms (quantum vs. classical, matter vs. spacetime, information vs. physical, observer vs. system) are dissolved into a single, relationally informed tapestry.

\section{Philosophical Implications and Commitments}
Relational Quantum Dynamics is not just a tweak to the physics formalism; it carries significant \textit{metaphysical and epistemological commitments}. Here we discuss how RQD relates to major positions in philosophy of physics and address some likely philosophical objections. We highlight the stance of RQD in terms of ontology (what exists), epistemology (how we know and describe) and its potential to bridge other debates (realism, emergence, mind–body, etc.).

\subsection{Ontology}
 \textit{Structural Realism and Informational Metaphysics:} RQD can be seen as an embodiment of \textit{ontic structural realism (OSR)} \cite{alai2017debates,allori2017scientific}. Instead of a world made of individual particles with intrinsic properties (as naive realism might hold), RQD suggests that the world is fundamentally \textit{relations of quantum information}. Objects such as electrons, detectors, or even spacetime points are secondary, emerging as relatively stable patterns within the network of relations. This aligns with the claim of OSR that 'there are no individuals, only relational structures' \cite{alai2017debates}. Ladyman and French have argued, using quantum theory’s puzzles, that if particles are indistinguishable and entangled with no independent identity, we might be forced to accept that \textit{relations are ontologically basic} and objects are derivative \cite{ladyman2007every,french2003remodelling}. RQD takes this notion literally: a quantum state is a state of affairs between systems, spacetime links are communication/entanglement relations, and even an observer is a higher-order relation (an integrated bundle of information). By removing intrinsic properties and basing the reality on a 'net of relations connecting all different physical systems', RQD's worldview resonates strongly with OSR and Whiteheadian \textit{process philosophy} \cite{whitehead1929process}, where reality is made up of events or interactions rather than substances. It also echoes John Wheeler’s famous slogan “It from Bit,” suggesting that information (the bit) underlies \textit{things} (it) \cite{Wheeler1989}.

However, RQD’s informational ontology goes further into what might be considered \textit{idealistic} territory, albeit with a twist. If one asks, “What is the stuff of the world in RQD?”, the answer could be: it is stuff \textit{made of information}, or even \textit{mind-like} in a certain sense \cite{Kastrup2017}. Since information by itself is an abstract notion, one might interpret RQD as positing something akin to a cosmic information field or “universal awareness”. Indeed, RQD implies a kind of \textit{monism}: there is one type of existence (information/relations) that, when organized one way, appears as the physical world and, when organized another way, constitutes conscious experience. This view bears similarity to some forms of neutral monism or panpsychism, where the basic constituents of reality have both physical and mental aspects. RQD is cautious in this claim: it does not say that electrons are conscious in any human-like way. Rather, it says that if you take information to be fundamental, then what we call “physical reality” and “conscious mind” are two manifestations of that information, from the outside and inside perspective respectively. In philosophical terms, RQD could be seen as a variant of \textit{idealistic structural realism} or \textit{information-theoretic ontology}: the world is a structure of information, and our minds are segments of that structure observing itself \cite{floridi2008defence,ladyman2007every,bynum2013quantum}.

Such a stance is admittedly provocative. It transgresses the usual realist vs. anti-realist dichotomy by introducing elements of idealism, information and perhaps consciousness as fundamental, yet insisting it is not subjective idealism or solipsism. RQD’s metaphysics might be described as \textit{participatory realism}: reality is there, but it inherently involves participatory relations. It asserts that observation is not a passive revelation of pre-set facts, but an active relation that \textit{constitutes} facts. This resonates with the \textit{participatory universe} envisioned by Wheeler and with Heisenberg’s idea that “what we observe is not nature itself, but nature exposed to our method of questioning.” In RQD, every “method of questioning” (interaction) is part of nature’s basic fabric.

\subsection{Epistemology}
 \textit{Perspectival Knowledge and Intersubjectivity.} Epistemologically, RQD implies that \textit{all descriptions are indexical}. Any statement about a quantum system such as “the cat is alive” or “the spin is up” is incomplete unless one specifies \textit{to whom} or \textit{relative to what basis} that statement applies. This is a radical departure from classical objectivity, where we imagine there’s a single true state of affairs that any competent observer would report. In RQD, by contrast, knowledge is \textit{perspectival}, which means different observers may have legitimately different knowledge states about the same system. This sounds similar to the doctrine of \textit{relationalism about truth} or \textit{perspectival realism} as discussed by philosophers like Simon Saunders or Michela Massimi (in general philosophy of science) \cite{massimi2022perspectival,saunders2003indiscernibles}. It also has parallels in the interpretation of quantum mechanics: QBism, for instance, says that a quantum state is not a element of reality but rather an agent’s information, and thus each agent has their own state assignment \cite{Fuchs2011-FUCAQR}. RQD agrees with QBism that state assignments are observer-dependent, but RQD \textit{maintains that this does not make physics purely subjective}. There \textit{is} an external reality, but it is such that any \textit{access} to it yields observer-bound facts. One might say RQD’s world has \textit{no “view from nowhere”}, only views from everywhere.

A critical epistemological issue is \textit{intersubjective agreement}: if every observer has their own truth, how do we explain the consensus reality we all seem to share? RQD addresses this by noting that when observers interact and exchange information, their previously separate relational descriptions \textit{become correlated and agree}. The key is that no observer is infallible or all-knowing; each has partial information. But when one observer measures another, or two observers compare notes, quantum theory (with decoherence in play) guarantees that relative records will align. This is essentially the \textit{“communication restores objectivity”} idea: while facts are not globally absolute, they can become effectively objective within any given reference frame that encompasses the communicating parties. Thus, RQD can recover the domain of classical intersubjective reality as an emergent phenomenon: we all agree on the objects and events in the room after we have interacted, because through interaction we become part of one joint quantum system with a common history. Before interaction, it was meaningless to compare our separate “stories”; after interaction, there is one story for the larger system that includes us both.

In philosophical terms, RQD suggests a reconciliation between the extreme relativism of “each has their own reality” and the need for scientific communicability. It requires refining what we mean by “fact”; a fact is always \textit{relative to a cognitive frame}, but facts are \textit{translatable} from one frame to another when a physical interaction connects those frames. One might see an analogy with relativity theory: simultaneity is relative, but different observers can relate their measurements via Lorentz transformations. In RQD, truth is relative, but observers can relate, and agree on, their truths via physical interaction (essentially via the unitary quantum dynamics that brings correlations). RQD thereby aims to avoid the pitfall of outright \textit{solipsism}. The world is not “just my imagination,” because there are consistency constraints on any one observer’s experiences and because any two observers can test and achieve agreement on overlapping aspects of the world once they interact. However, the \textit{limit} of this process is that we can never construct a single global view that is simultaneously valid for all; some perspectives like Wigner’s and the friend’s prior to interaction remain incommensurable. For a philosophy of science audience, this might echo Bohr’s notion of \textit{complementarity}, different experimental setups reveal different, mutually incompatible aspects of quantum systems. RQD generalizes complementarity to observers: different observers’ accounts are complementary until brought into interaction.

\subsection{Comparison with Other Interpretations}
 It is useful to explicitly contrast RQD with a few key interpretative stances:

\begin{itemize}
    \item \textbf{Vs. Copenhagen:} RQD agrees with Copenhagen that outcomes are contextual (depend on experimental setup) and that one must relinquish the idea of observer-independent properties. However, Copenhagen left a lurking classical observer outside the theory; RQD instead promotes \textit{everything} to a quantum description and finds the “classical” through emergent principles (decoherence, integrated information). So, RQD eliminates the Heisenberg cut by extending quantum ontology to observers, rather than positing an unanalyzed measurement axiom.
    \item \textbf{Vs. Many-Worlds (Everett):} RQD, like Everett, is fully quantum and eschews collapse. But Everett still posits a single universal wavefunction and treats all branches as equally “real” (simply non-interacting). RQD denies a single universal wavefunction in favor of many relational state-ascriptions. One could cheekily say RQD is a “many-realities interpretation”, but crucially these many realities are not parallel universes in the Everett sense; they are perspective-bound slices of one relational network. Everett’s theory has an absolutist meta-ontology (the wavefunction of the universe) with a multiplicity of emergent worlds, whereas RQD has no single absolute state at all, only perspectival states. Everett faces the “preferred basis” problem (what defines branches?) and “probability” problem (why the Born rule?). RQD addresses the former with decoherence (similar to Everett) and addresses the latter by noting that each observer sees Born-rule frequencies in their own frame by design, since RQD does not alter quantum mechanics’ predictions. In Everett, all outcomes happen but we perceive one because of decoherence splitting; in RQD, all outcomes \textit{relative to each decohered frame} exist, but there is no frame in which contradictory outcomes co-exist, so each observer just sees one.
    \item \textbf{Vs. Bohmian Mechanics:} Bohm’s pilot-wave theory retains a single reality (particles with positions) and is deterministic but nonlocal. RQD shares with Bohm the desire for a coherent ontology, no fundamental collapse or vagueness, but it does not introduce hidden variables or trajectories. Instead of restoring classical-like determinism, RQD accepts indeterminism (Born probabilities) as reflecting something real about information, perhaps relating to objective chance or propensities in the relational structure. Also, Bohmian mechanics requires a preferred frame (to define the pilot wave evolution) and absolute time, which RQD’s principles deliberately avoid. Bohm’s theory is explicitly \textit{non-relational}, it has an actual configuration for the whole universe. So philosophically, Bohmian mechanics is a kind of \textit{metaphysical Newtonianism}, particles in absolute space with a guiding wave. RQD is a kind of \textit{metaphysical relativism}, no absolute space, no absolute state – only relations.
    \item \textbf{Vs. QBism:} We touched on this earlier, QBism says quantum states and probabilities are an agent’s personal beliefs, and it emphasizes the subjective aspect of Bayesian updating upon measurement. RQD takes a lot of inspiration from QBism’s \textit{agent-centered perspective}, but RQD diverges by insisting that the relational quantum state is not \textit{merely} a belief, it is an \textit{element of reality for that agent}. In other words, RQD does not think the world is made of \textit{beliefs}; it is made of information-states that correspond to actual physical relations, which an agent might know imperfectly or partially. One could say RQD is \textit{QBism with ontology}. It tries to keep the empirical success of QBism (making sense of quantum collapse as belief-updating) while avoiding the implication that nothing objective exists. A slogan might be: \textit{the quantum state is an observer-dependent fact, not just an opinion.} We can also mention quantum \textit{perspectival realism} approaches (for example, views developing RQM further), RQD is very much in that camp, but it enriches it by including spacetime and other structures in the ontology.
    \item \textbf{Vs. RQM:} Whereas Relational Quantum Mechanics (RQM) insists that all quantum states are meaningful only \textit{relative} to an interacting system and denies any single, observer-independent wavefunction of the universe \cite{Rovelli1996}, Relational Quantum Dynamics (RQD) not only embraces this core insight but extends it in four crucial respects.  First, RQD internalizes the temporal parameter by invoking the \textit{thermal‐time hypothesis} \cite{connes1994von}, so that each subsystem’s quantum state generates its own flow of time, rather than presupposing an external clock.  Second, whereas RQM remains agnostic about the status of space and simply assumes a background arena, RQD proposes that \textit{spacetime itself is emergent} from patterns of quantum entanglement \cite{VanRaamsdonk2010}, with “distance” arising from the strength of informational links.  Third, RQDs supply a \textit{quantitative criterion}—drawn from Integrated Information Theory \cite{Tononi2016IntegratedInformationTheory}—to distinguish genuine “observers” (systems with high integrated information) from mere physical subsystems, thus formalizing a notion that in RQM is left informal.  Fourth, RQD embeds \textit{environment-induced decoherence} and \textit{Quantum Darwinism} \cite{zurek2009quantum,Zurek2005} into its ontology, showing how stable, redundant records proliferate in the environment so that different observers reliably agree on outcomes; RQM, by contrast, appeals only to abstract consistency of relational state assignments without specifying the dynamical mechanism for classical objectivity.  In addition to these advances, RQD draws on—and yet decisively differs from—other perspectival and topos‐based approaches, for example, Healey’s pragmatist quantum realism \cite{healey2012quantum}, Isham and Butterfield’s topos formulations \cite{Isham2000}, which reframe quantum logic or epistemic roles but do not integrate state-dependent time, entanglement-built geometry, information-based observer criteria, and decoherence-centered emergence into a single coherent framework.  By weaving these information-theoretic and spacetime-generating elements together, RQD goes well beyond RQM’s original ambition of resolving the measurement problem: it offers a unified ontology in which quantum mechanics, the emergence of space-time, and the physics of observers all flow from one relational, informational substrate.
\end{itemize}

\subsection{Mind–Body and Consciousness}
 One striking implication of RQD via the integrated information principle is a new perspective on the mind–body problem. If we accept that high integrated information indicates a system that “feels like one whole”, then RQD provides a natural way to say \textit{what consciousness is} in physical terms: it is the state of being a high-$\Phi$ quantum-information structure. This is in line with IIT’s claim that consciousness is “the capacity of a system to integrate information”. RQD suggests that a conscious observer (like a human brain) is simply a complex quantum system with a very large amount of internal correlation, enough to form a unified viewpoint. The \textit{dual-aspect} flavor is clear: from the outside, we describe the brain in RQD terms as a big integrated information node; from the inside, that integration \textit{is} what it feels like to be that mind. This view is akin to panpsychism in that it ascribes a sort of proto-experiential quality (integrated, holistic information) to matter, but it is also very restrictive: only systems that meet a high threshold of complexity have anything like what we’d call consciousness. It’s not that “electrons are conscious” – rather, electrons have essentially zero integrated information individually, so they don’t constitute observers on their own. RQD’s take is more like \textit{panprotopsychism} or \textit{informational monism}: fundamental reality is one kind (information) which, in certain structures, gives rise to consciousness. This has the benefit of avoiding the elusive interaction problem of dualism (mind and body aren’t separate substances if both are just information patterns), but it does verge into speculative metaphysics. We emphasize that RQD does not rely on any special intervention of consciousness to make the physics work (unlike the old “consciousness causes collapse” interpretations). Consciousness is along for the ride as a phenomenon that happens in complex systems, not a driver of quantum dynamics. In this sense, RQD is consonant with a scientifically grounded approach: it places conscious observers within the quantum world, subject to its laws, and even offers a quantitative handle (integrated info) to discuss them. This opens a dialogue with philosophy of mind: RQD hints at a solution to the hard problem by proposing that experience is what information structure “feels like from the inside”. Admittedly, this is an assertion rather than an explanation, but it aligns with certain philosophers (like Spinoza’s double-aspect idea, or modern proponents of integrated information like Tononi and Koch) in trying to marry the mental and physical in a single framework.

\subsection{Emergence vs. Reduction}
RQD is staunchly on the \textit{emergentist} side of the emergence/reduction debate. Classical reality (stable objects, spacetime continuum, definite outcomes) is emergent from the underlying quantum relational substrate. That means many properties at the classical level (for example, the definiteness of the cat’s alive/dead status, or the flow of time, or the geometry of a table) are \textit{novel features} that arise from complex interactions, not present in the micro-state of isolated components. RQD provides specific mechanisms for emergence: decoherence for classicality, entanglement for spatial connectivity, integrated information for unified systems, etc. These correspond to what philosophers might call \textit{strong emergence} in some cases, particularly integrated information, which is explicitly irreducible to local properties. Critics of strong emergence often worry it is incoherent or unscientific, but here it is grounded in precise theories (quantum mechanics and information theory). RQD therefore contributes to the debate by offering a concrete model where higher-level phenomena, such as the definiteness of an outcome or the presence of an “observer”, \textit{supervene} on but are not straightforwardly reducible to lower-level states. One can still, in principle, describe the entire world with the quantum formalism,which is reductionistic in spirit, but without the emergent principles one does not understand the appearance of classical reality. This is reminiscent of how thermodynamics emerges from statistical mechanics: no single molecule has a temperature, but a huge number of them collectively do. Likewise, no single qubit has a spacetime metric or a conscious experience, but a vast entangled ensemble might. Philosophically, RQD suggests that \textit{higher-level truths} like “the pointer read 5 at time $t$” are real and meaningful, though they are relative. They exist because of an emergent stable pattern within the relational dynamics. This viewpoint can enrich discussions about reduction: it exemplifies how a reductive \textit{description} (quantum wavefunctions) might fail to transparently display the ontology of the world, which only becomes clear when considering emergent, relational structures.

\subsection{Causation and Information:}
If information is fundamental in RQD, one may ask: how does causation work? Traditional physical causation might be described in terms of forces or energy exchange. In RQD, interactions are fundamentally \textit{information exchanges}. For instance, when two particles scatter, they become entangled, that is the exchange of information about each other’s states. This leans toward a concept of \textit{informational causation}, where what causes changes in one subsystem is the transfer or sharing of information from another. Some philosophers and physicists like John Wheeler have imagined that \textit{information flows} could be the most basic causal thread of reality \cite{wheeler1990information}. RQD gives this a specific meaning: a cause is an interaction that correlates the states of two systems, i.e., creates entanglement or decoheres a system with respect to an environment. As a result, “X caused Y” would translate to “the state of X became correlated with (and thus determined the state of) Y through an interaction”. This is compatible with physical causation but emphasizes relational, not intrinsic, terms. A particle does not cause another to move because of a force emanating from its mass alone, but because a field mediates information about one to the other as encoded in field quanta. This view dovetails with \textit{structural realist} accounts of causation that focus on patterns and relations rather than billiard-ball impacts.

One might worry that if the world is fundamentally information, do we risk saying nothing \textit{physical} at all causes events (since information is abstract)? RQD would respond that \textit{information is physical}. It is always instantiated in physical states (qubits, fields, what have you) and constrained by physical law. The difference is in emphasis: instead of thinking in terms of material substances exerting influence, we think of states carrying information and that information being propagated and transformed. This perspective could potentially illuminate long-standing issues such as the arrow of time (information entropy increases give direction to the arrow of time) or quantum causal structures (which in some advanced quantum information work even allow for indeterminate causal order). In RQD, because time is emergent and tied to entropy, \textit{causation might not be a fundamental linear order}, but something that itself emerges when a clear time parameter emerges. This is speculative, but it suggests RQD could engage with cutting-edge philosophical discussions on the nature of causality in a quantum world; for example, the work of Huw Price on retrocausality or discussions of causal sets in quantum gravity \cite{price2008toy,surya2019causal}. At minimum, RQD pushes us to conceive of causation in terms of information flow in a network, rather than the motion of independent objects.

\subsection{Objections and Replies:}
We now consider some likely objections to RQD and outline responses:

\textit{\textbf{Objection 1:} “This is just another interpretation, it does nor make new predictions. Isn't RQD subject to the same empirical underdetermination as any interpretation, and thus untestable?”} \textbf{Response:} It is true that RQD, as presented, is an \textit{interpretation} in the sense that it is built to recover all standard quantum predictions, so it does not blatantly contradict existing experiments. However, RQD’s merit is in providing a \textit{conceptual unification} that suggests new avenues for investigation. For instance, by linking entanglement to geometry, RQD implies that experiments at the interface of quantum information and gravity, for example, observing entanglement-related effects in spacetime geometry, perhaps in tabletop quantum gravity experiments, could support its view. One concrete example: if spacetime connectivity truly depends on entanglement, then perhaps one could detect gravitational effects disappearing when quantum entanglement is destroyed. There are proposals along these lines for witnessing gravitationally induced entanglement between masses as a clue that gravity is quantum. Similarly, the integrated information principle in RQD could, in principle, be tested by seeing if high-$\Phi$ systems behave differently at the quantum-classical boundary. For example, is there a threshold complexity beyond which superpositions become effectively impossible to maintain due to internal decoherence? This might be explored in mesoscopic quantum experiments. Moreover, RQD might guide new theory development. For example, a quantum gravity theory that naturally incorporates these principles, or a refinement of quantum theory’s formalism to explicitly include context-dependent state spaces (some work in algebraic quantum theory or topos theory aims at that). While empirical underdetermination is real (many interpretations fit the same data), RQD sets itself apart by straddling quantum foundations \textit{and} cosmology/quantum-gravity in a single narrative. It thus can be falsified in a broader sense if, say, the idea of emergent spacetime from entanglement fails, if instead spacetime is found to be fundamental down to the Planck scale, or if integrated information turns out to have no physical relevance to when quantum systems produce definite records. These are subtle, long-term testable implications. In the near term, RQD can be fruitful by suggesting thought experiments, for example, a “Wigner’s friend with entangled clocks” scenario to test relational time, that could reveal internal consistency or inconsistency in various interpretations. If RQD is conceptually inconsistent or leads to a contradiction with quantum probability in some scenario, that would be a way to falsify it. In short, while RQD shares the present experimental confirmation of standard QM with other interpretations, it is \textit{not} purely irrefutable: its strong claims about spacetime and information mean it could be disconfirmed by progress in quantum gravity or neuroscience. If, for instance, consciousness does not scale with integrated information at all, that would challenge pillar 4.

\textit{\textbf{Objection 2:} “By denying observer-independent reality, RQD sounds solipsistic or anti-realist. And by tying reality to information and even consciousness, isn’t this veering into unfalsifiable metaphysics or panpsychism?”} \textbf{Response:} RQD does take a bold stance that classical, observer-independent reality is an emergent approximation, not fundamental. This can be intellectually unsettling, but it is motivated by the concrete paradoxes of quantum theory (Wigner’s friend, Frauchiger–Renner) that seem to \textit{force} us to abandon the idea of a single absolute reality to avoid contradictions. RQD is \textit{realist} in the sense that it posits a mind-independent structure (the quantum-informational universe) that exists regardless of human observation. It is \textit{not} saying “nothing exists unless observed” in a subjective idealist way; it says \textbf{“what exists is a web of relations, and an ‘observer’ is just one part of that web relating to another part.”} This is a subtle but important distinction. It means RQD can still do physics objectively: the relations have mathematical descriptions and law-like dynamics. The \textit{solipsism concern} is addressed by showing how RQD allows for consistent communication of information between agents, yielding a shared effective reality. As we discussed, any theory giving up a single global reality must supply a mechanism for intersubjective agreement to avoid true solipsism. RQD provides that mechanism in quantum terms: interactions correlate different viewpoints so that they can be translated into each other. In practice, this means RQD’s predictions for everyday experiments are the same as usual quantum mechanics---scientists in a lab all agree on the outcomes after they compare notes, as normal. So one could live “as if” there is a single reality within any given lab frame; RQD simply cautions that extending that assumption outside the domain of interaction (to a superobserver who sees all) leads to paradox.

Regarding \textit{panpsychism/idealism}: RQD is careful not to assert that every electron or rock has consciousness. It uses the framework of integrated information to say that some physical systems (with very rich internal relational structure) instantiate something like mind, while others do not. This is more nuanced than traditional panpsychism, where every particle has a dash of consciousness. One might categorize RQD’s view as \textit{information monism} or \textit{dual-aspect monism}: information is the underlying stuff, and what we call physical or mental are two sides of how information organizes \cite{thompson2003problem,nemirovsky2023implementation}. Is this unfalsifiable? Possibly at the current stage, but it is an attempt to solve a puzzle (the observer’s role) without leaving physics. It draws on an existing, empirically grounded theory. IIT’s measures can be, and have been, applied to brain data, for instance \cite{toker2019information}. If integrated information turned out to be unrelated to consciousness in neuroscientific studies, that would undermine using it as a principle in RQD. Conversely, if future experiments show, say, that systems with low integrated information never seem to produce “classical” records (maybe highly dis-integrated systems remain quantum in ways integrated ones do not), that would support RQD’s stance. As for idealism: RQD’s emphasis on mind-like aspects of reality is speculative and not strictly required for its physical predictions; it is more a philosophical interpretation of what the mathematics might imply. \textbf{One could adopt RQD’s five principles in a deflationary way and just say “the world is quantum information” without commenting on consciousness at all.} Our inclusion of the IIT insight is meant to show one way to identify observers within the ontology, not to insist the universe is literally conscious. We also anchor these ideas by pointing to physics-friendly philosophies: for example, Wheeler’s participatory universe, or von Neumann–Wigner’s suggestion (though RQD modifies it significantly) that information and observation are fundamental in quantum theory. In short, RQD does flirt with philosophically adventurous ideas, but it remains \textit{constrained by physics}. It does not invoke anything outside the known quantum/information framework; it rather says perhaps physics \textit{itself} has a layer that could be called “mental” if seen from the right angle. Ultimately, whether one calls that panpsychism or just a radical kind of structural realism is a matter of preference. RQD’s test will be its utility: if it helps make sense of thorny problems and perhaps guides new insights, then its metaphysical boldness is a virtue. If it ends up adding no explanatory power, then it will join the heap of interpretations that are more philosophy than science. We have tried to ensure RQD \textit{earns} its keep by resolving known paradoxes and unifying disparate ideas.

\subsection{Broader Significance:}
If RQD (or something like it) is on the right track, it has implications beyond quantum physics. It suggests a new synthesis in our understanding of nature:

\begin{itemize}
    \item In the realm of \textbf{cosmology and quantum gravity}, RQD’s principles align with approaches where space and time are emergent (for example, holographic principle, AdS/CFT correspondence \cite{tHooft1993dimensional,maldacena1998large}), and where the quantum universe has no outside observer. It encourages physicists working on quantum gravity to incorporate quantum information as not just a calculational tool but as the substance of spacetime. This mirrors recent trends where entanglement entropy has geometric significance in gravity \cite{RyuTakayanagi2006}. RQD could thus be a philosophical cheerleader for programs like the \textbf{it-from-qubit} approach in quantum gravity.
    \item In \textbf{quantum information science}, RQD provides a narrative that the phenomena they study (entanglement, decoherence, information processing) are not just practical resources but the very fabric of reality. This might inspire new protocols or experiments. For instance, one might deliberately test the “entanglement creates geometry” idea in a simulation or look for a relation between integrated information and error-correcting codes in holography, since AdS/CFT hints that spacetime behaves like an error-correcting code.
    \item For \textbf{philosophy of science}, RQD exemplifies a move away from reductionist materialism to a more relational ontology. It resonates with \textbf{perspectival realism} \cite{massimi2022perspectival}, the idea that scientific truths are perspective-dependent but not merely subjective, and it provides a working example of how such a view can be made concrete. It also feeds into discussions of \textit{pluralism} in science: maybe there is not one true description of the world, but many partial ones that overlap. RQD shows how that can be the case without relativism going wild, because the partial views are related by precise transformation rules (quantum dynamics).
    \item In the \textbf{mind–body debate}, as discussed, RQD suggests the line between physical and mental is not a chasm but a spectrum of informational complexity. This might encourage a fresh look at panpsychist or neutral monist philosophies, grounding them in quantum concepts rather than 19th-century metaphysics. It also may give a framework to discuss \textit{downward causation}: if higher-level structures (like a mind) are real patterns of information, could they have causal efficacy? In RQD, higher-level patterns certainly influence lower-level dynamics (an observer’s integrated state can shape the decoherence environment, etc.) but ultimately everything is one network of unitary evolution. This parallels debates on whether emergent properties can feed back causally.
    \item Finally, RQD has a \textbf{participatory or pragmatic} side: it underscores that the role of the observer/agent is unavoidable in physics---not as a mysterious consciousness, but as a physical part of the system. This might foster more dialogue between physics and fields like \textit{philosophy of information} or even \textit{ethics}. Some have speculated if “participatory reality” has ethical implications \cite{hackett2013participatory}. For example, does the universe require observers to actualize it? RQD would say the universe \textit{is} observers relative to each other, which is a rather democratic vision of existence).
\end{itemize}

\section{Conclusion and Outlook}
\textbf{Philosophical Takeaway:} Relational Quantum Dynamics offers a bold re-imagining of quantum reality: one in which \textbf{quantum mechanics, spacetime, and observers are woven into a single ontological fabric}. The core message is that by relaxing certain classical assumptions, such as one absolute reality, a universal time, intrinsically separate objects, we can resolve long-standing paradoxes and unify insights from disparate domains of physics. Reality, in this view, is \textit{relational and informational}---a veritable \textbf{network of interacting perspectival realms}. What we perceive as a stable external world emerges from deeper quantum relations through mechanisms like decoherence and entanglement. Objectivity is recovered as an emergent, \textit{intersubjective agreement} rather than a fundamental God's-eye predicate. This represents a shift from the traditional Newtonian paradigm (of substance and absolute spacetime) to a \textit{Leibnizian/Machian paradigm} where relations and information are primary. In philosophical terms, RQD can be seen as advancing a form of \textbf{realism without absolutism}: it is realist in that it postulates a mind-independent structure (the quantum network governed by law), but it denies that any single observer or description captures that structure in its entirety---truth comes in perspectival layers.

\textbf{Future Directions in Physics:} To develop RQD further, one direction is to formalize these ideas within an existing or new theoretical framework. For instance, one could work on an extension of algebraic quantum field theory that incorporates contextual state assignments, each algebra of observables is indexed by an observer. Some work in topos theory and quantum logics by Isham and Butterfield hints at how to handle contextual truth values in quantum theory \cite{Isham2000,butterfield1999emergence}. In quantum gravity, exploring models of spacetime emergence via entanglement (already underway in AdS/CFT and tensor network approaches) will be crucial to put Principle 3 on firmer footing. If a precise “dictionary” between entanglement structure and spatial geometry can be established in broader contexts, it would strongly support RQD’s ontology. Experimentally, a daring but not unimaginable goal is an \textit{experimental test of relational quantum effects}. For example, a Wigner’s friend experiment is being pursued with small quantum systems, where one qubit “observes” another and a super-observer measures both. If such experiments demonstrate the consistency of different observers’ viewpoints only when quantum theory’s relational nature is accounted for, it could provide empirical backing for RQD’s interpretation. Another possible experiment: testing whether a system with variable integrated information (maybe a controllable network of qubits that can be tuned between isolated and highly entangled configurations) shows a threshold in behaving classically (does decoherence finalize outcomes faster or more robustly when $\Phi$ is high?).

\textbf{Future Directions in Philosophy:} On the philosophy side, RQD opens several avenues. It provides a rich case study for debates on \textit{scientific realism}: rather than arguing realism vs anti-realism, one can analyze RQD as an example of a third way (structures are real, but “facts” are relative). Philosophers can scrutinize whether RQD truly resolves the measurement problem or just shifts language---does relational truth satisfy our demand for an explanation of why I see one outcome? RQD says “because relative to you, only that outcome exists”---a satisfactory answer for some, but it invites probing of the nature of explanation itself. RQD also intersects with \textit{metaphysics of identity}: if objects have no identity outside relations, this resonates with bundle theory or structural realism debates on what individuates particles. The answer here being: nothing except their relations. There is also room to refine the \textit{concept of observer} in philosophy of physics: RQD suggests criteria (integrated information) for when something counts as an observer, which philosophers could compare to other criteria (like functionalist or cognitive definitions of observer). Additionally, RQD’s heavy use of information invites collaboration with philosophy of information---can we make precise the claim that information is the substance of reality? Is this a physically meaningful statement or a re-labeling of the quantum state concept? Clarifying these questions will sharpen RQD’s claims.

Finally, there are broader \textbf{humanistic implications} worth contemplating. If observers are part of reality’s fabric and not external, the strict Cartesian cut between subject and object weakens. Some might find in RQD a hint of overcoming the subject-object dichotomy that has arguably plagued Western thought. While we must be careful not to over-interpret, it is intriguing that a rigorously physical theory ends up with a picture of \textbf{“universe as a self-relational process”}---almost a poetic image, yet grounded in equations. This resonates with certain non-Western philosophies, for example, Buddhist and Vedantic ideas that the world is a net of interdependences and that the distinction between observer and observed is conventional. Such parallels should not be overstated, but they indicate RQD’s potential to foster dialogue between science and philosophy on questions of reality and experience.

\textbf{Concluding Remark:} Relational Quantum Dynamics is undoubtedly an ambitious framework, and many aspects remain to be fleshed out or tested. It may ultimately require refinement or even replacement as our empirical and theoretical understanding grows. Nonetheless, its significance lies in the integrative vision it offers: a single coherent story that ties together quantum measurement, the emergence of spacetime, and the nature of observers. The philosophical payoff is a possible resolution of deep tensions---between quantum and classical, between physics and information, between mind and matter---by recognizing that all of these dualities were pointing to the same fundamental insight: \textbf{reality is not made of things, but of relationships}. In embracing that insight, RQD provides a platform for both physicists and philosophers to jointly explore a world where what exists is not absolute “physical stuff” or disembodied information, but a cosmic conversation of information---a conversation in which we, as observers, are participants and not mere spectators. This, we contend, is a worldview capable of making sense of quantum mechanics \textit{and} of ourselves within a single ontological picture, something that a century of discussions has struggled to achieve. The work is far from complete, but RQD charts a path toward a more unified understanding of reality, inviting further inquiry, criticism, and refinement from the community of physicists and philosophers alike.

\textbf{Acknowledgement:}
The core concepts, theoretical constructs, and novel arguments presented in this article are a synthesis and concretization of my own original ideas. At the same time, in the process of assembling, interpreting, and contextualizing the relevant literature, I used OpenAI's GPT as a tool to help organize, clarify, and refine my understanding of existing research. In addition, I utilized OpenAI reasoning models and sought their assistance in refining the mathematical derivations. The use of this technology was instrumental for efficiently navigating the broad and often intricate body of work in quantum theory, category theory, and IIT. 

\bibliography{references} 

\end{document}